\documentclass[12pt]{iopart}
\usepackage{xcolor}
\usepackage{graphicx}

\begin{document}

\title{Electrostatic potential and electric field in the $z$ axis of a non centered circular charged ring}

\author{F. Escalante}

\address{Departamento de Física, Universidad Católica del Norte, Avenida Angamos 0610, Casilla 1280, Antofagasta, Chile.}
\ead{fescalante@ucn.cl}
\vspace{10pt}
\begin{indented}
\item[]August 2021
\end{indented}

\begin{abstract}
In introductory level electromagnetism courses the calculation of electrostatic potential and electric field in an arbitrary point is a very common exercise. One of the most viewed cases is the calculation of electrostatic potential and electric field in the symmetry axis of a centered ring and it has been widely studied the potential off the axis of a charged ring centered in the origin coordinate. In this work, we calculated the electrostatic potential and electric field in the $z$ axis of a non centered charged ring using elliptic integrals as an pedagogical example of the application of special functions in electromagnetism.
\end{abstract}

\maketitle
\section{Introduction}
The calculation of electrostatic potential at an arbitrary point, due to a charge distribution, is a general topic presented to undergraduate students at introductory electromagnetism course~\cite{serway, sears, tipler}. Taking advantage of the scalar nature of potential it can calculate the Electric field at an arbitrary point. Typical examples are the calculation of the electrostatic potential of a sphere, a long rod in an arbitrary point, as well as a disk and uniformly charged ring, over a point of his symmetry axes. Nevertheless, in the two last cases when we want to calculate the electrostatic potential at any point of the space off-axis, due to this charge distribution, we must deal with special functions as elliptic integrals~\cite{good, ciftja, noh, ciftja2,Bochko} or Legendre polynomials~\cite{jackson}.
In this article, we calculate the electrostatic potential of a non-centered charged ring in the $z$ axis using a suitable parametrization of a circle of radius $R$.  This result allows recovering the well-known expressions for the potential of a centered ring on its symmetry axes and outside of it.

\section{Electrostatic potential on the axis}
The electrostatic potential in terms of a charge density is given by~\cite{jackson}

\begin{equation}
    \label{potencial}
    \Phi(\vec{r})=\frac{1}{4\pi\epsilon_0}\int \frac{\rho(\vec{r}^{\,'})}{\mid \vec{r}-\vec{r}^{\,'}\mid}d^{3}\vec{r}^{\,'}.
\end{equation}

\begin{figure}[h]
    \centering
    \includegraphics[width=0.4\linewidth]{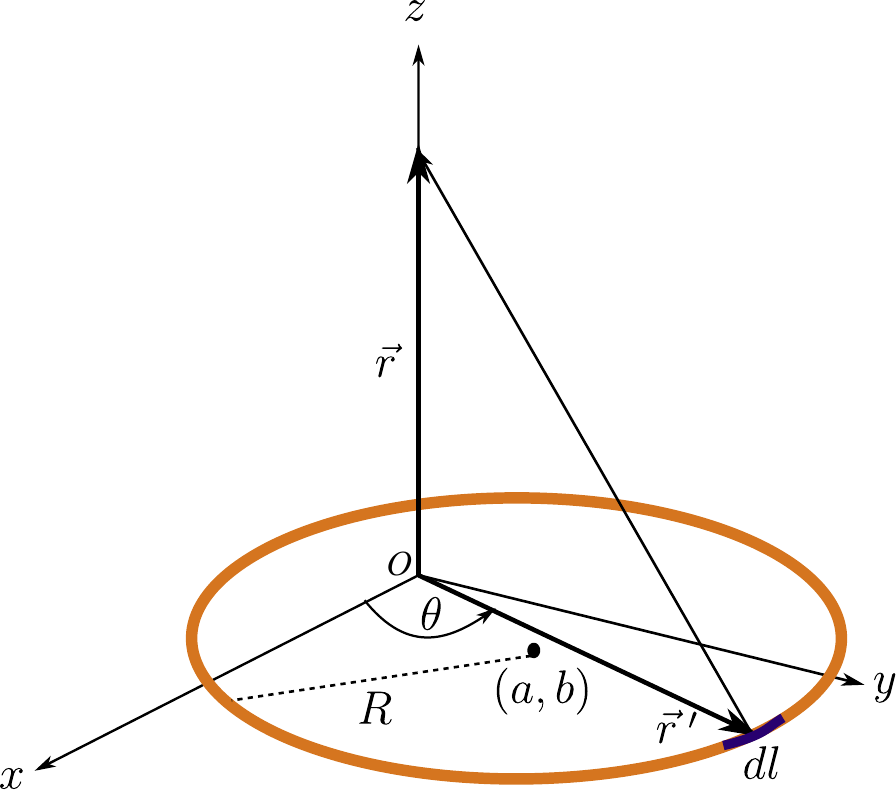}
    \caption{Graphic representation of the considered situation.}
    \label{f1}
\end{figure}

We consider a uniformly charged ring of radius $R$ and uniform charge density $\lambda=Q/2\pi R$ located in the $xy$ plane which is non centered respect to the origin coordinate $O$ as shown in figure (\ref{f1}) 

\noindent so we have 

\begin{equation}
    \label{pot1}
    \Phi(\vec{r})=\frac{\lambda}{4\pi\epsilon_0}\int\frac{dl}{\mid \vec{r}-\vec{r}^{\;'}\mid}.
\end{equation}

Using the parametric equation of a non centered ring given by $x'=a+R\cos\theta$ and $y'=b+R\sin\theta$, where ($a$,$b$) corresponds to the coordinates of the center of the ring and $0\leq\theta\leq 2\pi$. Therefore $\vec{r}=z\hat{k}$ and $\vec{r}^{\,'}=(a+R\cos\theta)\hat{\imath}+(b+R\sin\theta)\hat{\jmath}$, so we can obtain the infinitesimal length element by means of 

\begin{equation}
    \label{dl}
    dl=\mid d\vec{r}^{\,'}\mid=\sqrt{(dx')^2+(dy')^2}=Rd\theta.
\end{equation}

Then, the electrostatic potential in the $z$ axis is given by

\begin{equation}
    \label{pot2}
    \Phi(z)=\frac{\lambda}{4\pi\epsilon_0}\int_{0}^{2\pi}\frac{R\,d\theta}{\sqrt{a^2+b^2+R^2+z^2+2aR\cos\theta+2bR\sin\theta}}
\end{equation}    

\noindent defining $\alpha=a/R$, $\beta=b/R$, $\zeta=z/R$ and $\Lambda=\alpha^2+\beta^2+\zeta^2+1$ we have

\begin{equation}
\label{potn}
    \Phi(\zeta)=\frac{\lambda}{4\pi\epsilon_0}\int_{0}^{2\pi}\frac{d\theta}{\sqrt{\Lambda+2\alpha\cos\theta+2\beta\sin\theta}},
\end{equation}

The integral in the equation (\ref{potn}) can be written as~\cite{gradshteyn}

\begin{equation}
\label{int1}
   I= \int_{0}^{2\pi}\frac{d\theta}{\sqrt{\Lambda+2\alpha\cos\theta+2\beta\sin\theta}}=\int_{-\varphi}^{2\pi-\varphi}\frac{du}{\sqrt{\Lambda+\varepsilon\cos u}}.
\end{equation}

The parameters of equation (\ref{int1}) are defined by $\varphi=\arctan(\beta/\alpha)$ and $\varepsilon=2\beta/\sin\varphi=2\sqrt{\alpha^2+\beta^2}$ and the variable $u=\theta-\varphi$. Using the transformation $u=2\phi$ we can re write the denominator of integrand (\ref{int1}) as

\begin{equation}
    \label{}
    (\Lambda+\varepsilon\cos u)^{1/2}=(\Lambda+\varepsilon)^{1/2}\left(1-m\sin^{2}\phi\right)^{1/2},
\end{equation}

\noindent where $m=2\varepsilon/(\Lambda+\varepsilon)$. So the equation (\ref{int1}) can be written as

\begin{equation}
\label{I}
    I=\frac{2}{\sqrt{\Lambda+\varepsilon}}\int_{-\varphi/2}^{\pi-\varphi/2}\frac{d\phi}{\sqrt{1-m\sin^{2}\phi}}.
\end{equation}

The solution of the integral (\ref{I}) is given by the incomplete elliptic integral of the first kind $F(\varphi,m)$~\cite{AYS}  

\begin{equation}
F(\varphi,m)=\int_{0}^{\varphi}(1-m\sin^{2}\phi)^{-1/2}d\phi,
\end{equation}

\noindent then, the electrostatic potential in $z$ axis is

\begin{equation}
    \label{pot3}
    \Phi(\zeta)=\frac{\lambda}{2\pi\epsilon_0}\frac{1}{\sqrt{(1+\gamma)^2+\zeta^2}} \left[F\left(\frac{\varphi}{2},\frac{4\gamma}{(1+\gamma)^2+\zeta^2}\right)-
      F\left(-\pi+\frac{\varphi}{2},\frac{4\gamma}{(1+\gamma)^2+\zeta^2}\right)\right]
\end{equation}

\noindent where $\gamma=\sqrt{\alpha^2+\beta^2}$.

Since $F(0,m)=0$ and $F(-\pi,m)=-2K(m)$, where $K(m)$ is the complete elliptic integral of the first kind in terms of the parameter $m$, we can make $\beta=0$ and $\alpha=0$ then $m=0$. Noticing that $K(0)=\pi/2$, we have

\begin{equation}
    \label{potcent}
    \Phi(z)=\frac{Q}{4\pi\epsilon_0}\frac{1}{\sqrt{R^2+z^2}},
\end{equation}

\noindent which is the electrostatic potential in the $z$ axis of a centered ring. 
\begin{figure}
    \centering
    \includegraphics[width=0.6\textwidth]{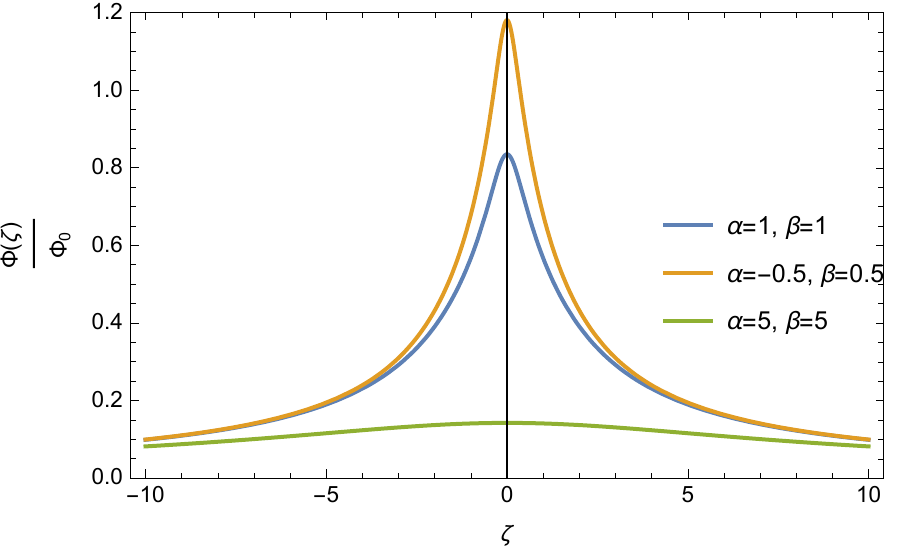}
    \caption{Electrostatic potential $\Phi(z)$ (eq.~\ref{pot3}) normalized to the potential in $\Phi(0)=\Phi_0$ of a unitary radius ring for different positions of the ring in the plane.}
    \label{f2}
\end{figure}

As we can see from the figure \ref{f2} the normalized electrostatic potential if the center of the ring moves away form the origin coordinate, the normalized electrostatic potential in the $z$ axis tends to vanish.

If we compare this situation with the electrostatic potential, in the space, of a centered ring~\cite{good, ciftja, noh}, we can consider that the ring its displaced a distance $\gamma=\sqrt{\alpha^2+\beta^2}$ from the origin coordinate $O$. Considering $\gamma$
to lie on the plane of the ring along the $x$ direction, then $\beta=0$ so $\gamma=\alpha=\varrho$, where $\varrho=\rho/R$. Therefore, the electrostatic potential is given by

\begin{equation}
\label{potpart}
 \Phi(\varrho,\zeta)=\frac{\lambda}{2\pi\epsilon_0}\frac{2}{\sqrt{(\varrho+1)^2+\zeta^2}}K\left[\frac{4\varrho}{(1+\varrho)^2+\zeta^2}\right],
 \end{equation}
 
 \noindent or in terms of $z$
 
 \begin{equation}
 \Phi(\rho,z)=\frac{Q}{4\pi^2\epsilon_0}\frac{2}{\sqrt{(\rho+R)^2+z^2}}K\left[\frac{4R\rho}{(\rho+R)^2+z^2}\right].
\end{equation}

\section{Electric field}

As is known, the electric field can be obtained from the potential gradient. In cylindrical coordinates, considering cylindrical symmetry, the electric field is given by

\begin{equation}
\vec{E}=-\vec{\nabla}\Phi(\rho,z)=-\left(\frac{\partial}{\partial \rho}\hat{\rho}+\frac{\partial}{\partial z}\hat{k}\right)\Phi(\rho,z),
\end{equation}

\noindent however, in order to obtain an expression for the electric field in terms of the coordinates $(\varrho,\zeta)$, we must consider the change of variable $\varrho=\rho/R$ and $\zeta=z/R$ to preserve the dimensions of the electric field, so we have

\begin{equation}
\label{grad}
\vec{E}=-\frac{1}{R}\left(\frac{\partial}{\partial \varrho}\hat{\varrho}+\frac{\partial}{\partial \zeta}\hat{k}\right)\Phi(\varrho,\zeta).
\end{equation}

As we seen before, the expressions (\ref{pot3}) and (\ref{potpart}) are analogous, so we can obtain the $(\varrho,\zeta)$ components of the electric field by means of the gradient of equation (\ref{potpart}), using the expression (\ref{grad}).

Considering that

\begin{equation}
    \label{difk}
 \frac{dK(k)}{dk}=\frac{E(k)}{k(1-k^2)}-\frac{K(k)}{k}
\end{equation}

\noindent and

\begin{equation}
    \label{dife}
   \frac{dE(k)}{dk}=\frac{E(k)-K(k)}{k},
\end{equation}

\noindent where $E(k)$ is the incomplete elliptic integral of the second kind. After some manipulations, the radial component of the electric field is given by 

\begin{eqnarray}
\label{campor}
  E_{\varrho}=&\frac{\lambda}{2\pi\epsilon_{0}R}\frac{1}{\varrho\sqrt{(1+\varrho)^2+\zeta^2}((\varrho-1)^2+\zeta^2)}\nonumber\\&\times\left\{(\varrho^2-1-\zeta^2)E\left[\frac{4\varrho}{(1+\varrho)^2+\zeta^2}\right]+((1-\varrho)^2+\zeta^2)K\left[\frac{4\varrho}{(1+\varrho)^2+\zeta^2}\right]\right\},\nonumber\\
\end{eqnarray}

\noindent and the vertical component of the electric field is

\begin{equation}
    \label{campoz}
E_{\zeta}=\frac{\lambda}{2\pi\epsilon_{0}R}\frac{2\zeta}{\sqrt{(1+\varrho)^2+\zeta^2}((\varrho-1)^2+\zeta^2)}E\left[\frac{4\varrho}{(1+\varrho)^2+\zeta^2}\right].
\end{equation}

From the equations (\ref{campor}) and (\ref{campoz}) we can check the cases when $\varrho\to 0$. In this cases we must recover the expression for a centered uniformly charged ring for the vertical component and zero for the radial component of the electric field.

As we can see, the expression (\ref{campor}) can not be evaluated directly in $\varrho=0$ as it presents a $0/0$ behaviour. So we must calculate the limit as $\varrho\to 0$.

Considering (\ref{difk}) and (\ref{dife}), and after some manipulation we have

\begin{eqnarray}
\label{camporr}
E_{\varrho}=&\frac{\lambda}{2\pi\epsilon_{0}R}\frac{1}{\sqrt{(1+\varrho)^2+\zeta^2}(\varrho^3+4\varrho^4+5\varrho^2(\zeta^2-1)-\varrho(1+\zeta^2)+(1+\zeta^2)^2)}\nonumber\\
    &\times\left\{((1+\varrho)^3+(1+3\varrho)\zeta^2)E\left[\frac{4\varrho}{(1+\varrho)^2+\zeta^2}\right]\right.\nonumber\\
    &\left.+(\varrho-1)(1+\varrho(4+3\varrho)+\zeta^2)K\left[\frac{4\varrho}{(1+\varrho)^2+\zeta^2}\right]\right\}.
\end{eqnarray}
 
Evaluating the limit $\varrho\to 0$ in the expression (\ref{camporr}) and noticing that $E(0)=\pi/2$ and $K(0)=\pi/2$, we ended up with

\[
E_{\varrho}=0.
\]

On the other hand, we can calculate directly the electric field in the vertical component for $\varrho=0$ in the expression (\ref{campoz})

\begin{equation}
    \label{campozz}
 E_{\zeta}=\frac{\lambda}{2\epsilon_{0}R}\frac{\zeta}{(1+\zeta^2)^{3/2}}, 
\end{equation}
 
\noindent as $\zeta=z/R$ and $\lambda=Q/2\pi R$ we obtain the well known expression for the electric field of an centered uniformly charged ring along the symmetry axis

\begin{equation}
    E_{z}=\frac{Q}{4\pi\epsilon_{0}}\frac{z}{(R^2+z^2)^{3/2}}.
\end{equation}

\section{Conclusions \label{sec:concl}}

It has been presented a general method to obtain the electrostatic potential using an adequate parametrization of a loop through
the use of elliptic integrals. From this result, it could recover the expressions of the electrostatic potential and electric field on the symmetry axis of a centered ring. Finally, the obtained results agree with the calculations of the electrostatic potential off the symmetry axis of a centered ring and are useful as a pedagogical exercise for to students can explore special functions in an electromagnetism problem.

\section*{Acknowledgment}
The author thanks Professors Julio M Yáñez of Universidad Católica del Norte (Chile) and Juan P Ramos of Universidad Técnica Federico Santa María (Chile) for suggestions and time spent in many discussions.


\end{document}